# Anomalous spin-orbit field via Rashba-Edelstein effect at W/Pt interface

Shutaro Karube[1,2], Nobuki Tezuka[1,2], Makoto Kohda[1,2,3] and Junsaku Nitta[1,2,3]

[1] *Department of Materials Science, Graduate School of Engineering, Tohoku University, Sendai, Japan*
[2] *Center for Spintronics Research Network, Tohoku University, Sendai, Japan*
[3] *Center for Science and Innovation in Spintronics (Core Research Cluster) Organization for Advanced Studies, Tohoku University, Sendai, Japan*

**Abstract**

We have studied spin-orbit (SO) field in $Ni_{80}Fe_{20}$(Py)/W/Pt trilayer by means of spin-torque ferromagnetic resonance, and demonstrated that the W/Pt interface generates an extra SO field acting on the Py layer. This unprecedented field originates from the following three processes, 1) spin accumulation at W/Pt interface via the Rashba-Edelstein effect, 2) diffusive spin transport in the W layer, and 3) spin absorption into the Py layer through accumulation at the Py/W interface. Our result means that we can create extra SO field away from the ferromagnet/ metal interface and control its strength by a combination of two different metals.

Spin-orbit (SO) field or torque generated by spin Hall effect (SHE) [1] from SO materials can effectively perform magnetization switching of an adjacent ferromagnetic layer [2-5]. The SO field is based on a spin current density $\mathbf{J_S^{int}}$ acting on a ferromagnet (FM) interface shown in Eq. (1) [6,7].

$$\mathbf{J_S^{int}} = \mathrm{Re} G_{\uparrow\downarrow}(\mathbf{m} \times \boldsymbol{\mu} \times \mathbf{m}) + \mathrm{Im} G_{\uparrow\downarrow}(\mathbf{m} \times \boldsymbol{\mu}) \qquad (1)$$

Here $G_{\uparrow\downarrow}$, $\boldsymbol{\mu}$, $\mathbf{m}$ are spin mixing conductance at the FM interface, the vector of spin accumulation, and the unit vector of magnetization in the FM layer, respectively. This accumulation $\boldsymbol{\mu}$ is necessary for generating the SO torque on the magnetization $\mathbf{m}$. The first and second terms in Eq. (1) correspond to damping-like (DL) and field-like (FL) torques, respectively. If the accumulation $\boldsymbol{\mu}$ is modulated, we can manipulate the magnitude of the SO torques.

Previously heavy metals such as Pt, Ta, and W with strong SO coupling have been intensively studied on various aspects, e.g. the SO torque induced switching, domain wall displacement [8], spin relaxation [9,10], and controllable SO torque [11,12] for fundamental physics and spintronics applications. Moreover it was found that surface states of topological insulators and Rashba states at interfaces can also generate SO torque or spin current via the Edelstein effect [13-20], resulting in effective magnetization switching because of their giant SO torque [21-23]. Thus, the magnitude of the SO torque in various materials have been systematically investigated and unveiled. For further efficiency of the SO torque or extra SO torques, novel SO materials are desired both in research and in practical applications.

Recently, there have been several new approaches in synthesizing novel SO materials. One such approach is by using oxides such as $CuO_X$ [24,25], $WO_X$ [26], and $PtO_X$ [27,28]. These materials' spin-orbit torque generations are believed to originate from the Rashba-Edelstein (RE) effect. Through the introduction of oxygen, there can be an enhancement of the torque compared to the pure metal, or the torque can be controlled via ionic-oxygen conduction by gating. Another approach is an SO heterojunction consisting of heavy metals (HMs), such as in sandwich structures [29] and bilayers [30]. Concerning sandwich structures (HM1/FM/HM2), the generated SO torque would be enhanced due to double spin accumulation at the top and bottom FM interfaces originating from the heavy metals, which have opposite signs of the spin Hall angle relative to each other, for example Pt and Ta [31,32]. These reports suggest that we can design new SO materials by such combinations. In the case of bilayers, generated SO torque has succeeded in performing field-free switching in W/Pt/CoFeB/MgO structures [30]. Thus the challenges to synthesize SO materials would be an important role for spintronics in the future. Based on previous reports regarding the RE effect [15,20,33], we have the possibility to generate extra SO field in the ferromagnet far from the HM interface with diffusive spin transport in the HM layer. In this study, we demonstrate an additional SO-field from the W/Pt interface.

We prepared the multilayer films by means of DC and RF magnetron sputtering at room temperature with reference samples as follows: $Ni_{80}Fe_{20}$(Py, 5)/W(2)/AlO$_X$ (2), Py(5)/W(1-5)/Pt(1), Py(5)/W(2)/Pt(1-4), Py(4-7)/W(2)/AlO$_X$(2), Py(4-7)/Pt(1), Py(5)/Pt(1)/AlO$_X$(2), Py(5)/Pt(1)/W(2)/Mg(1.5) (thickness in nanometers). Here AlO$_X$, and Mg which was naturally oxidized were employed as capping layers for the prevention of oxidation. According to the X-ray diffraction measurements, all metallic films are polycrystalline. Next, devices for spin-torque ferromagnetic resonance (ST-FMR) [34] were fabricated by conventional lift-off processes using e-beam lithography and Ar ion milling. For the ST-FMR measurement, we applied radio-frequency (RF) currents in the range of 3 to 14 dBm-power and 5 to 12 GHz-frequency from a signal generator into the devices, and detected DC voltages via anisotropic magnetoresistance (AMR) rectification while applying a static magnetic field $H$ up to 290 mT.

First of all, we demonstrate the generation of the extra SO field in the Py layer via the RE effect at the W/Pt interface. A schematic image is shown for the stacking structure and the spin current generation in Fig. 1 (a). Figure 1 (b) shows the detected DL fields $H_{DL}$ divided by the applied current density $J_C$ in the case of Py(5)/W(2)/AlO$_X$(2), Py(5)/Pt(1), and Py(5)/W(2)/Pt(1) at 6 dBm and 8 GHz of RF current using the below Eq. (2) [25]. The detected voltage can be described as $V_{mix} = V_S L(H) + V_{AS} \partial L(H)/\partial H$, with the Lorentzian $L(H)$ as a function of the magnetic field $H$, the amplitude of the Lorentzian $V_S$, and the amplitude of the derivative Lorentzian $V_{AS}$, and cannot be used for comparison due to different values of several parameters such as the applied RF current $I_{rf}$, AMR amplitude $\Delta R$, and the half width at half maximum of the ST-FMR spectrum $\mu_0 \Delta H$ in each sample.

$$\mu_0 H_{DL} = \frac{2V_S}{I_{rf}\Delta R} \frac{2\sqrt{2}\pi f \mu_0 \Delta H(2\mu_0 H_R + \mu_0 M_{eff})}{\gamma(\mu_0 H_R + \mu_0 M_{eff})\mu_0 H_R} \tag{2}$$

Here $\gamma, f, \mu_0 H_R, \mu_0 M_{eff}$ are gyromagnetic ratio for electron, applied frequency, resonance field and effective magnetization. As for the Py/W and the Py/Pt cases, the signs of the detected DL fields are in good agreement with previous reports [4,31] for both cases. In our case of 1nm-thick Pt, there is insufficient spin relaxation in the Pt layer according to the spin diffusion length of Pt: $\lambda_{sf}^{Pt} = 1.0 \pm 0.1$ nm which we measured, meaning the $V_{mix}$ amplitude is smaller than that which is originated from the saturated spin Hall angle in the case of a Pt layer thicker than $\lambda_{sf}^{Pt}$. So if there is only bulk contribution to the generation of the DL field, we should see a summation between W and Pt which includes a decay of the spin current in the W layer because of spin relaxation. Since the DL field from 1 nm-thick Pt layer is quite small, then the DL field in the case of the Py/W/Pt trilayer should be approximately the same as the Py/W case. However, the amplitude for Py/W/Pt seems to be reduced. This implies that there is another contribution generating the extra SO field as shown in the red curve in Fig. 1 (b). Moreover we measured a control sample consisting of Py(5)/W(2)/Cu(5)/Pt(1) multilayer with Cu insertion between the W and Pt layers in Fig. 1(c). The extra component disappears in this case, indicating the importance of the W/Pt interface. This behavior is similar to another report regarding the disappearance of the SO fields by Cu insertion between a ferromagnet and Pt [35]. We also checked the contribution of the Pt/AlOx interface via the RE effect as shown in Fig. 2(a) by comparing it to the Pt-only sample, and found that there is no difference between them, and so no contribution from the Pt/AlOx interface. Thus we have investigated the DL fields generated from the W and Pt layers via the spin Hall effect, and a negligible field from the Pt surface (or the Pt/AlOx interface) via the RE effect. Therefore, this extra field should emerge from the W/Pt interface via the RE effect, because it is not possible to explain from the other contributions.

To confirm the origin of the extra SO field arising from the RE effect at the W/Pt interface, we focus on another SO field, i.e. the FL field. We investigated the ST-FMR signal for the cases of Py(5)/W(2)/AlOx(2) and Py(5)/W(2)/Pt(1) at 6 dBm and 8 GHz of RF current, and surprisingly detected almost symmetric voltages in both cases, as seen in Fig. 2(b). To clarify these symmetric voltages, we investigated only Pt(1) or W(2)/AlOx(2), while varying the Py thickness, and extracted the DL and FL spin torque efficiencies $\xi_{DL}$ and $\xi_{FL}$ respectively, based on a reference of [36]. The efficiencies were calculated to be $\xi_{DL}^{Pt} = 0.0244$, $\xi_{FL}^{Pt} = -0.0016$, $\xi_{DL}^{W} = -0.0427$, $\xi_{FL}^{W} = 0.0167$ in the Py/W and Py/Pt systems. Concerning the 1nm-thick Pt case, the positive $\xi_{DL}^{Pt}$, which is related to the effective spin Hall angle, is in good agreement with a reported value [34] with a consideration for the spin relaxation in Pt as $\xi_{DL} = \xi_{DL}^0[1 - \text{sech}(t_{HM}/\lambda_{sf}^{HM})]$ [37], because this Pt thickness is the same as the spin diffusion length of Pt: $\lambda_{sf}^{Pt} = 1.0 \pm 0.1$ nm, which we measured. Negligible $\xi_{FL}^{Pt}$ is also consistent with the previous report [34]. Concerning the 2 nm-thick W case, the negative sign of $\xi_{DL}^{W}$ is consistent and the amplitude is close to a reported value for $\alpha$-W [4]. The finite $\xi_{FL}^{W}$ is in good agreement with the value for W/CoFeB system [38]. From this analysis, the almost symmetric voltage in the Py(5)/W(2)AlO$_X$(2) case is attributed to the finite FL field which is comparable to Oersted field from the W layer, but with the opposite sign, because the anti-symmetric voltage $V_{AS}$ consists of a summation between the Oersted field and the FL field shown below in Eq. (4).

Next, we have to consider why we observed almost symmetric voltages in the ST-FMR measurement of Py(5)/W(2)/Pt(1) as well. The amplitudes of the symmetric and anti-symmetric voltages are related to the SO fields including the DL, FL, and Oersted fields which are described in Eqs. (3) and (4) [36].

$$V_S \propto H_{DL} = \frac{\hbar}{2e} \frac{\xi_{DL} J_C}{\mu_0 M_S t_{FM}} \tag{3}$$

$$V_{AS} \propto (H_{FL} + H_{Oe})\sqrt{1 + \frac{\mu_0 M_{eff}}{\mu_0 H_R}} = \left(\frac{\hbar}{2e}\frac{\xi_{FL} J_C}{\mu_0 M_S t_{FM}} + \frac{J_C t_{HM}}{2}\right)\sqrt{1 + \frac{\mu_0 M_{eff}}{\mu_0 H_R}} \tag{4}$$

Here $e, \hbar, H_{DL}, H_{FL}, H_{Oe}$, and $J_C$ represent elementary charge, Dirac constant, DL field, FL field, Oersted field, and charge current density, respectively. $V_{AS}$ is almost negligible, meaning the $H_{FL}$ in the Py(5 nm)/W(2 nm)/Pt(1 nm) system has a comparable amplitude and opposite sign as the $H_{Oe}$ from the W and Pt layers. In other words, $H_{FL}^{tot} = -(H_{Oe}^W + H_{Oe}^{Pt})$. This compensation is not actually strange because we have already seen a similar kind of signal in the Py(5 nm)/W(2 nm)/AlO$_X$(2 nm), where comparable FL and Oersted fields emerge at the Py/W interface, i.e. $H_{FL}^W = -H_{Oe}^W$. If the Pt layer, instead of the AlO$_X$ layer, creates an additional Oersted field in the Py layer, then the almost symmetric voltage should be broken and an anti-symmetric voltage should appear. But, the detected voltage is still symmetric. As we pointed out before, this is due to the contribution of the W/Pt interface for generating an extra effective FL field, i.e. $H_{FL}^{W/Pt} \approx -H_{Oe}^{Pt}$.

One possible doubt is that these results are coming from a spin pumping effect during the FMR [39], which could explain the almost symmetric voltage in ST-FMR measurement. To rule this spin pumping effect out, we also measured the in-plane field-angular dependence of the signals and RF power dependence as shown in Figs. 2 (c) and (d), and found $\sin 2\theta_H \cos \theta_H$ behavior which is attributed to the ST-FMR signal [40], and a linear relationship between the detected voltage and the power. This ensures that the ST-FMR measurements and analysis were performed precisely, and reliable.

Now we must consider how both the extra DL and FL fields can be created via the W/Pt interface. As stated previously, we propose that these results originate from the RE effect which generates spin accumulation at the W/Pt interface. The RE effect has already been demonstrated in several systems such as in the surface states of topological insulators [14,15] and Rashba interfaces [17-20,33,41-45] which have spin-split dispersion at Fermi level. Therefore, the non-equilibrium accumulated spins propagate with decay along the out-of-plane direction (through the W and Pt layers) diffusively. If the W thickness $t_W$ is smaller or comparable to the spin diffusion length of W: $\lambda_{sf}^W \approx 1.1$ nm as we extracted, and in good agreement with other reports [46,47], some portion of the accumulation can pass through the W layer, and finally reach at the Py/W interface. Then, the spin absorption and accumulation take place, and the extra DL and FL fields emerge into the Py layer, separate from the SHE contribution.

Based on this assumption, we now focus on quantitative analysis to extract the anomalous SO fields. First of all, we calculated charge current densities and Oersted fields in and from each layer considering shunting effects and applied power of the RF current. The injected RF current into the devices depends on the strip impedance. So an applied 6 dBm power is actually reduced by following RF current power relationship $P = (V/(R_I + R_S))^2 R_S$. Where $P, V, R_I$, and $R_S$ represent actual injected power, voltage, internal impedance of the signal generator, and impedance of the sample strip, respectively. The Oersted field from each layer can be simply estimated by $H_{Oe}^{HM} = J_C^{HM} t_{HM}/2$, introduced from the Gauss's law in the case of an infinite conductive plane. The fields divided by the charge current densities are shown in Fig. 3 (a). We also estimated the conventional DL and FL fields from the bulk W and Pt layers by following Eq. (5) [36] with extracted $\xi_{DL}^{HM}$ and $\xi_{FL}^{HM}$ values based on the argument above as shown in Figs. 3 (a) and (b).

$$H_{DL(FL)} = \frac{\hbar}{2e} \frac{J_C^{HM}}{\mu_0 M_S t_F} \xi_{DL(FL)} \qquad (5)$$

For the estimation of the DL field from the W layer, we used $\xi_{DL}^W = \xi_{DL}^{W,0}[1 - \mathrm{sech}(t_W/\lambda_{sf}^W)]$ based on the spin diffusion model with $\lambda_{sf}^W \approx 1.1$nm as obtained by us for the estimation. Moreover, for the DL field from the Pt layer, it has to propagate into W layer with spin diffusion. Therefore, we assumed a simple exponential decay [20,33] for the amplitude of the DL field, $H_{DL}^{Pt} = H_{DL}^{Pt,0} \exp(-t_W/\lambda_{sf}^W)$. Here $H_{DL}^{Pt,0}$ stands for the DL field generated in the 1nm-thick Pt layer before spin diffusion into the W layer. With these parameters, we can estimate the extra field contributions coming from the W/Pt interface by subtracting from the detected signal. According to Eq. (4), there is a cancellation of the fields between the Py(5)/W(2) and Py(5)/W(2)/Pt(1) cases as described in Eq. (6).

$$H_{\text{FL}}^{\text{W/Pt}} \exp(-t_{\text{W}}/\lambda_{\text{sf}}^{\text{W}}) = -H_{\text{Oe}}^{\text{Pt}} \tag{6}$$

Here, $H_{\text{FL}}^{\text{W/Pt}}$ means the FL field generated by the W/Pt interface and is related to the spin accumulation at the interface. By taking the value of the Oersted field from Pt layer, the normalized FL field $H_{\text{FL}}^{\text{W/Pt}}$ divided by the charge current density in the W layer is $H_{\text{FL}}^{\text{W/Pt}} \exp(-t_{\text{W}}/\lambda_{\text{sf}}^{\text{W}}) /J_{\text{C}}^{\text{W}} = 0.034 \text{ mT}/ 10^{11} \text{A m}^{-2}$ for the 2nm-thick W case. Here, the reason why we chose $J_{\text{C}}^{\text{W}}$ as the charge current density for the calculation is that the W layer is dominant for the conduction close to the W/Pt interface because of no evidence of interfacial alloying according to the phase diagram between W and Pt [48] and the energy difference between surface binding energy of the W: 10 eV [49] and the conventional sputtering energy: 2 eV [50]. The estimated FL field is comparable to the conventional FL field generated at the Py/W interface which has the normalized FL field as $H_{\text{FL}}^{\text{W}}/J_{\text{C}} = 0.138 \text{ mT}/ 10^{11} \text{A m}^{-2}$ in this study. We also estimated the DL field $H_{\text{DL}}^{\text{W/Pt}}$ from the W/Pt interface by following Eq. (7) based on Eq. (2).

$$\mu_0 H_{\text{DL}}^{\text{W/Pt}} \exp(-t_{\text{W}}/\lambda_{\text{sf}}^{\text{W}}) = \frac{2V_{\text{S}}}{I_{\text{rf}}\Delta R} \frac{2\sqrt{2}\pi f \mu_0 \Delta H (2\mu_0 H_{\text{R}} + \mu_0 M_{\text{eff}})}{\gamma(\mu_0 H_{\text{R}} + \mu_0 M_{\text{eff}})\mu_0 H_{\text{R}}} - \mu_0 H_{\text{DL}}^{\text{W}} - \mu_0 H_{\text{DL}}^{\text{Pt}} \exp(-t_{\text{W}}/\lambda_{\text{sf}}^{\text{W}}) \tag{7}$$

We now know the amplitudes of the DL fields from both W and Pt according to the calculation above. Therefore, it is possible to estimate the extra contribution from the W/Pt interface as shown in Fig. 3 (b). Here, we also assume simple exponential decay [20,33], as we did with the DL field from the Pt layer, and almost current going through the W layer. Surprisingly, we found that the amplitude of the DL field seen in Fig. 3 (b) is not negligible, but is comparable to the conventional DL field from the W layer in the case of thinner W thicknesses such as 1 nm. This means that we can create extra SO fields in the FM layer away from the HM interface. Here Fig. 3 (c) shows a ratio of the effective fields derived from the $V_{\text{S}}/V_{\text{AS}}$ raw data by using Eq. (8) [27] for comparison with the ratio calculated by a summation of extracted SO fields from the assumptions above.

$$\frac{H_{\text{Oe}} + H_{\text{FL}}}{H_{\text{DL}}} = \frac{V_{\text{AS}}}{V_{\text{S}}} \left(1 + \frac{\mu_0 M_{\text{eff}}}{\mu_0 H_{\text{R}}}\right)^{-1/2} \tag{8}$$

In spite of the presence of multiple SO field parameters, the ratio roughly follows the same behavior with respect to W thickness as the ratio from $V_S/V_{AS}$, implying that our calculation is a nearly correct estimation. As for the minimum position at 2 nm, it is attributed to the spin diffusion length (1.1 nm) of the W layer. The spin accumulation generated via the spin Hall effect in heavy metals or the RE effect at the W/Pt interface regarding the SO fields diffuses inside the W layer with decay. So overall, the minimum appears at around 2 nm. Moreover we also investigated the Pt thickness dependence of the extra DL field as shown in Fig. 4. Here we included the Pt thickness dependence of the DL torque efficiency of the Pt layer $\xi_{DL}^{Pt}$ for the extra DL field calculation. As can be seen, it decays with increasing thickness. This decay length was roughly estimated as $\lambda_{decay} = 2.0 \pm 0.7$ nm which is nearly the estimated spin diffusion length of the Pt ($\lambda_{sf}^{Pt} = 1.0 \pm 0.1$ nm). This behavior is in good agreement with our assumption in this study that the generated spins from the RE effect at the W/Pt interface diffuse into not only the W layer, but also the Pt layer. So, the amount of spins is reduced with increasing Pt thickness due to spin diffusion in the Pt layer.

Finally, we focus on an inverse stacking for W and Pt, by changing from Py/W/Pt to Py/Pt/W trilayer. If the assumption in which the Rashba effect at W/Pt interface contributes to the extra SO fields is true, we should see a difference in the $V_{mix}$ signal because the inverse stacking structure changes the direction of the effective electric field, and thus a different direction of the spin-polarization is also caused via the Rashba effect as shown in Fig. 5 (a). We would expect to see the opposite sign of the additional FL field in the case of Py/Pt/W as compared to Py/W/Pt. Thus we also performed the measurement for Py/Pt/W in the same thickness conditions, and we show the experimental results for both cases in Fig. 5 (b). As can be seen, the signal for the inverse stacking has an anti-symmetric voltage, implying that there was a large change of the SO fields from the previous stacking case of Py/W/Pt. For a detailed calculation of the FL field $H_{FL}^{Pt/W}$ from the Pt/W interface in the case of a Py/Pt/W trilayer system, we employed the following Eq. (9) based on reference [25].

$$\mu_0 H_{FL}^{Pt/W} \exp(-t_{Pt}/\lambda_{sf}^{Pt}) = \frac{2V_{AS}}{I_{rf}\Delta R} \frac{\sqrt{2}\mu_0 \Delta H(2\mu_0 H_R + \mu_0 M_{eff})}{\mu_0 H_R + \mu_0 M_{eff}} - \mu_0 H_{Oe}^{Pt} - \mu_0 H_{Oe}^{W} \quad (9)$$

As mentioned, we already know the amplitudes of the Oersted fields from the W and Pt layers. So, we can estimate the amplitude of $H_{FL}^{Pt/W}$ with subtraction, where we note that $H_{FL}^{Pt}$ is negligible based on the estimated quite small $\xi_{FL}^{Pt}$. The value of $H_{FL}^{Pt/W}$ divided by charge current density in the W layer is calculated to be $-0.0267$ (mT/$10^{11}$ A/m$^2$) for the 2 nm-thick W case. To estimate the field at the Pt/W interface, we have to consider exponential decay in the Pt layer, which has 1.0 nm-spin diffusion length which we estimated, in the same way as the above discussion. This amplitude was estimated to be $-0.0730$ (mT/$10^{11}$ A/m$^2$). Taking into account the decay in the W layer, which has 1.1 nm spin diffusion length we extracted, in the case of Py/W/Pt the amplitude of $H_{\text{FL}}^{\text{W/Pt}}/J_{\text{C}}^{\text{W}}$ is 0.3170 (mT/$10^{11}$ A/m$^2$).

We succeeded in finding the opposite sign of the extra FL field between Py/W/Pt and Py/Pt/W systems due to the sign change of the Rashba parameter through the inverse stacking. However, the value for the Py/Pt/W case was 4 times smaller than that for the Py/W/Pt case. The reason for this is related to the FL torque efficiency at the Py/HM interface. Py/W and Py/Pt interfaces have finite and negligible values of the $\xi_{\text{FL}}^{\text{HM}}$, respectively. This implies that there is a difference in the ability to accumulate spins at the interface between both cases. The Py/Pt interface cannot produce spin accumulation as effectively as Py/W. Thus, we do not see the same amplitude of the additional FL field. These experimental results ensure that this anomalous behavior is coming from the RE effect at the W/Pt interface and the diffusive spin transport process in the HM layers. Concerning further interfacial systems, we have to consider and calculate the electron density distribution and nuclei electric fields for amplitude of the Rashba parameter [33]. Moreover, we have to seek the same sign of the spin polarization between the spin Hall effect from heavy metal layers and the RE effect. We expect that there are good combinations because of the many reported SO materials.

In conclusion, we studied the Py/W/Pt trilayer system by means of ST-FMR with comparison to several reference samples carefully, and demonstrated the generation of extra SO fields via the RE effect at the W/Pt interface. This study concludes that we can create additional SO fields in the FM layer and control the amplitudes by a combination of HM layers which include an interface far from the FM layer remotely, and pave the way for more efficient and functional spintronic devices.

### Acknowledgement

We thank J. Ryu, H. Gamou, R. Thompson, R. Ando, and S. Fujikawa for their helpful information and constructive discussions. This work is partially supported by Japan Society for the Promotion of Science (JSPS) (Grants No. 15H05699, No. 17H06512, and No. 18K14111), Center for Spintronics Research Network in Tohoku University and Center for Science and Innovation in Spintronics in Tohoku University.

**References**

[1]     M. I. Dyakonov and V. I. Perel, JETP Lett. **13**, 467 (1971).

[2]     L. Liu, C.-F. Pai, Y. Li, H. W. Tseng, D. C. Ralph, and R. A. Buhrman, Science **336**, 555 (2012).

[3]     L. Liu, O. J. Lee, T. J. Gudmundsen, D. C. Ralph, and R. A. Buhrman, Phys. Rev. Lett. **109**, 096602 (2012).

[4]     C.-F. Pai, L. Liu, Y. Li, H. W. Tseng, D. C. Ralph, and R. A. Buhrman, Appl. Phys. Lett. **101**, 122404 (2012).

[5]     Q. Hao and G. Xiao, Phys. Rev. Applied **3**, 034009 (2015).

[6]     Y.-T. Chen, S. Takahashi, H. Nakayama, M. Althammer, S. T. B. Goennenwein, E. Saitoh, and G. E. W. Bauer, Phys. Rev. B **87**, 144411 (2013).

[7]     A. Brataas, Y. V. Nazarov, and G. E. W. Bauer, Phys. Rev. Lett. **84**, 2481 (2000).

[8]     S. Emori, U. Bauer, S. M. Ahn, E. Martinez, and G. S. Beach, Nat. Mater. **12**, 611 (2013).

[9]     J. Ryu, M. Kohda, and J. Nitta, Phys. Rev. Lett. **116**, 256802 (2016).

[10]    H. Gamou, J. Ryu, M. Kohda, and J. Nitta, Appl. Phys. Express **10**, 023003 (2017).

[11]    E. Lesne, Y. Fu, S. Oyarzun, J. C. Rojas-Sanchez, D. C. Vaz, H. Naganuma, G. Sicoli, J. P. Attane, M. Jamet, E. Jacquet, J. M. George, A. Barthelemy, H. Jaffres, A. Fert, M. Bibes, and L. Vila, Nat. Mater. **15**, 1261 (2016).

[12]    R. Mishra, F. Mahfouzi, D. Kumar, K. Cai, M. Chen, X. Qiu, N. Kioussis, and H. Yang, Nat. Commun. **10**, 248 (2019).

[13]    V. M. Edelstein, Solid State Commun. **73**, 233 (1990).

[14]    A. R. Mellnik, J. S. Lee, A. Richardella, J. L. Grab, P. J. Mintun, M. H. Fischer, A. Vaezi, A. Manchon, E. A. Kim, N. Samarth, and D. C. Ralph, Nature **511**, 449 (2014).

[15]    K. Kondou, R. Yoshimi, A. Tsukazaki, Y. Fukuma, J. Matsuno, K. S. Takahashi, M. Kawasaki, Y. Tokura, and Y. Otani, Nat. Phys. **12**, 1027 (2016).

[16]    Y. Wang, P. Deorani, K. Banerjee, N. Koirala, M. Brahlek, S. Oh, and H. Yang, Phys. Rev. Lett. **114**, 257202 (2015).

[17]    J. C. Sanchez, L. Vila, G. Desfonds, S. Gambarelli, J. P. Attane, J. M. De Teresa, C. Magen, and A. Fert, Nat. Commun. **4**, 2944 (2013).

[18]    A. Nomura, T. Tashiro, H. Nakayama, and K. Ando, Appl. Phys. Lett. **106**, 212403 (2015).

[19]    M. B. Jungfleisch, W. Zhang, J. Sklenar, W. Jiang, J. E. Pearson, J. B. Ketterson, and A. Hoffmann, Phys. Rev. B **93**, 224419 (2016).

[20]    S. Karube, K. Kondou, and Y. Otani, Appl. Phys. Express **9**, 033001 (2016).

[21]    Y. Fan, P. Upadhyaya, X. Kou, M. Lang, S. Takei, Z. Wang, J. Tang, L. He, L. T.

Chang, M. Montazeri, G. Yu, W. Jiang, T. Nie, R. N. Schwartz, Y. Tserkovnyak, and K. L. Wang, Nat. Mater. **13**, 699 (2014).

[22]    J. Han, A. Richardella, S. A. Siddiqui, J. Finley, N. Samarth, and L. Liu, Phys. Rev. Lett. **119**, 077702 (2017).

[23]    K. Yasuda, A. Tsukazaki, R. Yoshimi, K. Kondou, K. S. Takahashi, Y. Otani, M. Kawasaki, and Y. Tokura, Phys. Rev. Lett. **119**, 137204 (2017).

[24]    H. An, Y. Kageyama, Y. Kanno, N. Enishi, and K. Ando, Nat. Commun. **7**, 13069 (2016).

[25]    T. Gao, A. Qaiumzadeh, H. An, A. Musha, Y. Kageyama, J. Shi, and K. Ando, Phys. Rev. Lett. **121**, 017202 (2018).

[26]    K. U. Demasius, T. Phung, W. Zhang, B. P. Hughes, S. H. Yang, A. Kellock, W. Han, A. Pushp, and S. S. Parkin, Nat. Commun. **7**, 10644 (2016).

[27]    H. An, T. Ohno, Y. Kanno, Y. Kageyama, Y. Monnai, H. Maki, J. Shi, and K. Ando, Sci. Adv. **4**, eaar2250 (2018).

[28]    H. An, Y. Kanno, A. Asami, and K. Ando, Phys. Rev. B **98**, 014401 (2018).

[29]    J. Yu, X. Qiu, W. Legrand, and H. Yang, Appl. Phys. Lett. **109**, 042403 (2016).

[30]    Q. Ma, Y. Li, D. B. Gopman, Y. P. Kabanov, R. D. Shull, and C. L. Chien, Phys. Rev. Lett. **120**, 117703 (2018).

[31]    M. Morota, Y. Niimi, K. Ohnishi, D. H. Wei, T. Tanaka, H. Kontani, T. Kimura, and Y. Otani, Phys. Rev. B **83**, 174405 (2011).

[32]    H. L. Wang, C. H. Du, Y. Pu, R. Adur, P. C. Hammel, and F. Y. Yang, Phys. Rev. Lett. **112**, 197201 (2014).

[33]    H. Tsai, S. Karube, K. Kondou, N. Yamaguchi, F. Ishii, and Y. Otani, Sci. Rep. **8**, 5564 (2018).

[34]    L. Liu, T. Moriyama, D. C. Ralph, and R. A. Buhrman, Phys. Rev. Lett. **106**, 036601 (2011).

[35]    V. Ostwal, A. Penumatcha, Y.-M. Hung, A. D. Kent, and J. Appenzeller, J. Appl. Phys. **122**, 213905 (2017).

[36]    C.-F. Pai, Y. Ou, L. H. Vilela-Leão, D. C. Ralph, and R. A. Buhrman, Phys. Rev. B **92**, 064426 (2015).

[37]    M. H. Nguyen, D. C. Ralph, and R. A. Buhrman, Phys. Rev. Lett. **116**, 126601 (2016).

[38]    Y. Takeuchi, C. Zhang, A. Okada, H. Sato, S. Fukami, and H. Ohno, Appl. Phys. Lett. **112**, 192408 (2018).

[39]    K. Kondou, H. Sukegawa, S. Kasai, S. Mitani, Y. Niimi, and Y. Otani, Applied Physics Express **9**, 023002 (2016).

[40]    D. Fang, H. Kurebayashi, J. Wunderlich, K. Vyborny, L. P. Zarbo, R. P. Campion, A.

Casiraghi, B. L. Gallagher, T. Jungwirth, and A. J. Ferguson, Nat Nanotechnol **6**, 413 (2011).

[41]　J. Kim, Y.-T. Chen, S. Karube, S. Takahashi, K. Kondou, G. Tatara, and Y. Otani, Phys. Rev. B **96**, 140409(R) (2017).

[42]　J. Puebla, F. Auvray, M. Xu, B. Rana, A. Albouy, H. Tsai, K. Kondou, G. Tatara, and Y. Otani, Appl. Phys. Lett. **111**, 092402 (2017).

[43]　F. Auvray, J. Puebla, M. Xu, B. Rana, D. Hashizume, and Y. Otani, J. Mater. Sci.: Mater. Electron. **29**, 15664 (2018).

[44]　M. Xu, J. Puebla, F. Auvray, B. Rana, K. Kondou, and Y. Otani, Phys. Rev. B **97**, 180301(R) (2018).

[45]　K. Kondou, H. Tsai, H. Isshiki, and Y. Otani, APL Mater. **6**, 101105 (2018).

[46]　J. Kim, P. Sheng, S. Takahashi, S. Mitani, and M. Hayashi, Phys. Rev. Lett. **116**, 097201 (2016).

[47]　T.-C. Wang, T.-Y. Chen, C.-T. Wu, H.-W. Yen, and C.-F. Pai, Phys. Rev. Mater. **2**, 014403 (2018).

[48]　A. G. Knapton, Platinum Metals Rev. **24**, 64 (1980).

[49]　M. Gyoeroek, A. Kaiser, I. Sukuba, J. Urban, K. Hermansson, and M. Probst, J. Nucl. Mater. **472**, 76 (2016).

[50]　R. V. Stuart and G. K. Wehner, J. Appl. Phys. **35**, 1819 (1964).

Fig. 1

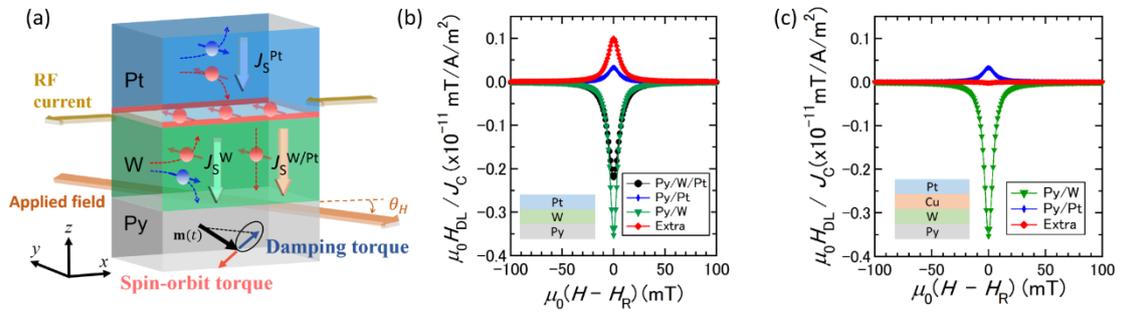

Fig. 1 (a) Schematic image showing each spin current $J_S$ generation via the spin Hall effect from the W and Pt layers and the Rashba-Edelstein effect from the W/Pt interface. During the FMR, a magnetic field is applied at the angle $q_H$ to the strip, and the magnetization of the Py layer feels the spin-orbit torque. (b) The generated damping-like field divided by charge current density $J_C$ for Py(5 nm)/W(2 nm) (green triangles), Py(5 nm)/Pt(1 nm) (blue diamonds), and Py(5 nm)/W(2 nm)/Pt(1 nm) (black circles). Red diamonds correspond to the extra component of the DL field. (c) The generated damping-like field divided by charge current density in the case of Py(5 nm)/W(2 nm)/Cu(5 nm)/Pt(1 nm).

Fig. 2

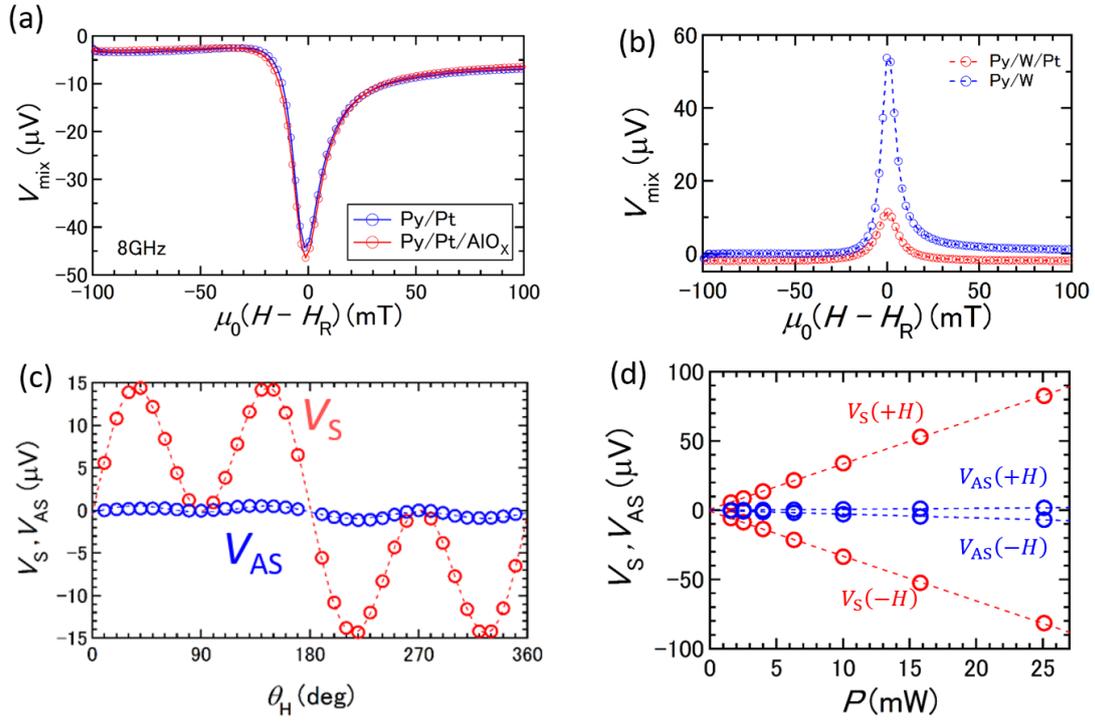

Fig. 2 (a) Detected voltages in the ST-FMR measurement at 8 GHz for Py(5 nm)/Pt(1 nm) (blue open circles) and Py(5 nm)/Pt(1 nm)/AlO$_X$(2 nm) (red open circles) with fitted curves. (b) Detected voltages in ST-FMR measurement at 8 GHz for Py(5 nm)/W(2 nm)/AlO$_X$(2 nm) (blue open circles) and Py(5 nm)/W(2 nm)/Pt(1 nm) (red open circles). (c) Field-angular dependence of the symmetric (red open circles) and anti-symmetric (blue open circles) voltages, and (d) RF power dependence of both of the voltages as a function of the applied field in the case of Py(5 nm)/W(2 nm)/Pt(1 nm).

Fig. 3

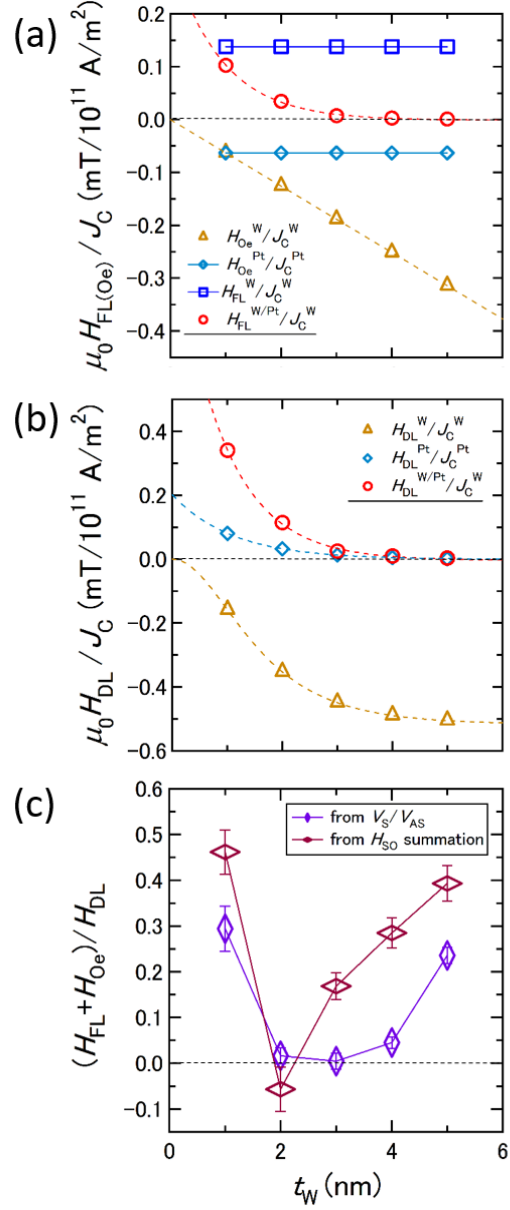

Fig. 3 (a) Estimated Oersted and FL fields divided by charge current density $J_C$ as a function of the W thickness in the cases of W(2 nm) bulk, Pt(1 nm) bulk, Py(5 nm)/W(1-5 nm) interface, and W(1-5 nm)/Pt(1 nm) interface with spin diffusion in W layer cases. (b) Estimated DL field divided by charge current density $J_C$ as a function of the W thickness in the cases of Pt(1 nm) bulk, W(2 nm)/Pt(1 nm) interface with spin diffusion in the W layer, and W(1-5 nm) bulk. (c) Comparison of field ratio $(H_{FL}+H_{Oe})/H_{DL}$ as a function of W thickness for estimation from raw $V_S/V_{AS}$ data and from $H_{SO}$ field summation.

Fig. 4

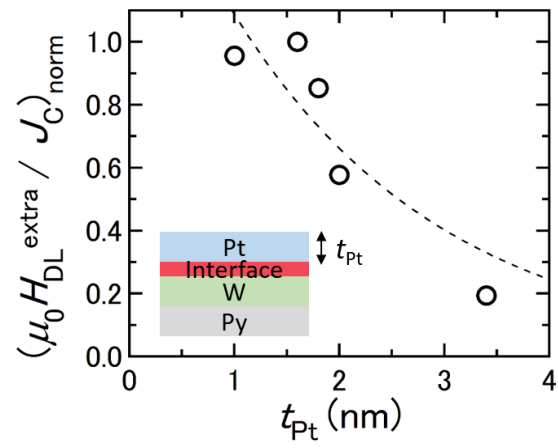

Fig. 4 Normalized damping-like field divided by charge current density as a function of the Pt thickness.

Fig. 5

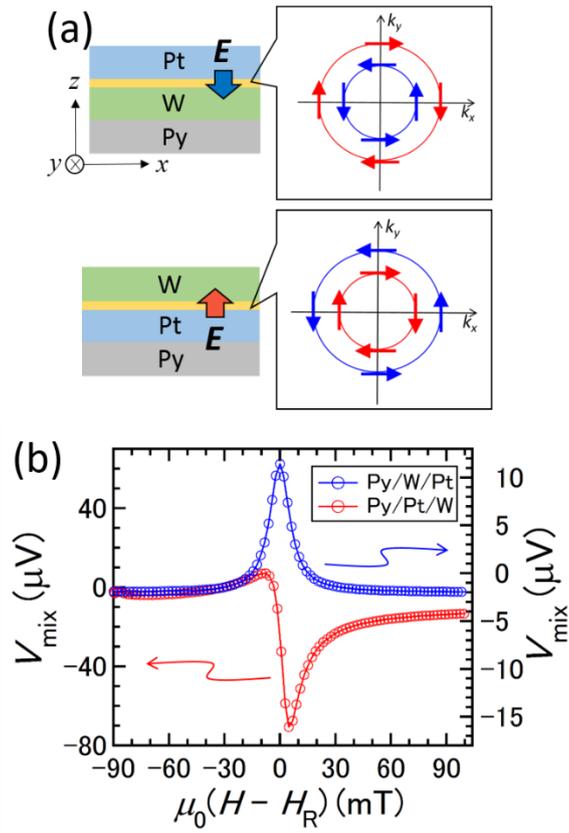

Fig. 5 (a) Cross sectional views of the structures with effective electric field $E$ at the heavy metals interface and expected Rashba spin polarization in cases of the normal stacking (Py/W/Pt) and the inverse stacking (Py/Pt/W). (b) Detected voltages in the ST-FMR measurement for Py(5 nm)/W(2 nm)/Pt(1 nm) and Py(5 nm)/Pt(1 nm)/W(2 nm).